\documentclass[12pt]{article}
\usepackage{amssymb,epsfig,times,graphicx}
\usepackage{amsmath}
\usepackage{citesort}
\usepackage{graphicx}
\numberwithin{equation}{section}

\newtheorem{thm}{Theorem}

\newtheorem{prop}{Proposition}

\newtheorem{lemma}{Lemma}

\newtheorem{rmk}{Remark}

\newcommand{\beq}{\begin{equation}}
\newcommand{\eeq}{\end{equation}}
\newcommand{\de}{\partial}
\def\d{\partial}
\def\n{\noindent}
\def\f{\frac}
\def\sa{\sum_\alpha\epsilon_\alpha}

\newcommand{\lm}{\lambda}

\newcommand{\pf}{\noindent{\it Proof \ }}

\newcommand{\epf}{\begin{flushright}
$\Box$\end{flushright}}
\begin{document}
 
\title{Purely nonlocal Hamiltonian formalism for systems of hydrodynamic type}

\author{John Gibbons${}^*$, Paolo Lorenzoni${}^{**}$, Andrea Raimondo${}^{*}$\\
\\
{\small * Department of Mathematics, Imperial College}\\
{\small 180 Queen's Gate, London SW7 2AZ, UK}\\
{\small j.gibbons@imperial.ac.uk, a.raimondo@imperial.ac.uk}\\
\\
{\small ** Dipartimento di Matematica e Applicazioni}\\
{\small Universit\`a di Milano-Bicocca}\\
{\small Via Roberto Cozzi 53, I-20125 Milano, Italy}\\
{\small paolo.lorenzoni@unimib.it}}
 
\date{}

\maketitle

\begin{abstract}
We study purely nonlocal
Hamiltonian structures for systems of hydrodynamic type.
In the case of a semi-Hamiltonian system, we show that such 
structures are related to quadratic expansions of the diagonal metrics
naturally associated with the system.   
\end{abstract}

\section*{Introduction}
In the last three decades many papers have been devoted 
to Hamiltonian structures for systems 
of hydrodynamic type:
\beq\label{hts} 
u^i_t=v^i_j(u)u^j_x,\qquad i=1,\dots,n.
\eeq 
The starting point of this research was the paper \cite{dn84}
(see also \cite{dn89}) where Dubrovin and Novikov introduced an 
important class of local Hamiltonian structures, called 
\emph{Hamiltonian structures of hydrodynamic type}. 
Such operators have the form
\begin{equation}\label{PBHT}
P^{ij}=g^{ij}(u)\frac{d}{dx}-g^{is}\Gamma^{j}_{sk}(u)u^k_x,
\end{equation}
where $g^{ij}$ are the contravariant components of a flat 
pseudo-Riemannian metric and  $\Gamma^j_{sk}$ are the Christoffel 
symbols of the associated Levi-Civita connection. Nonlocal extensions of the bracket (\ref{PBHT}), related to metrics
of constant curvature, were considered by Ferapontov and Mokhov in
\cite{fm90}; further generalizations,  of the form
\beq\label{ferop}
P^{ij}=g^{ij}\frac{d}{dx}-g^{is}\Gamma^j_{sk}u^k_x+\sum_{\alpha}\varepsilon_{\alpha}\left(w_\alpha\right)^i_ku^k_x
\left(\frac{d}{dx}\right)^{\!-1}\!\!\!\left(w_\alpha\right)^j_hu^h_x\,,\qquad\varepsilon_{\alpha}=\pm 1,
\eeq
were considered by Ferapontov in \cite{fe91}. Here we have defined
$$\left(\frac{d}{dx}\right)^{\!-1}=\frac{1}{2}\int^x_{-\infty}dx-\frac{1}{2}\int_x^{+\infty}dx,$$
and the index $\alpha$ can take values over a 
finite, infinite or even continuous 
set. In the case $\det{g^{ij}}\ne 0$, 
the operator (\ref{ferop})
defines a Poisson structure if and only if the tensor $g^{ij}$ defines 
a pseudo-Riemannian metric, the coefficients $\Gamma^j_{sk}$ are the 
Christoffel symbols of its Levi-Civita connection $\nabla$, 
and the affinors $w_\alpha$ satisfy the conditions 
\begin{subequations}\label{gmc}
\begin{gather} 
\left[w_\alpha,w_{\alpha'}\right]=0,\label{GMC1}\\
g_{ik}(w_{\alpha})^k_j=g_{jk}(w_{\alpha})^k_i,\label{GMC2}\\
\nabla_k(w_{\alpha})^i_j=\nabla_j(w_{\alpha})^i_k,\label{GMC3}\\
R^{ij}_{kh}=\sum_\alpha\varepsilon_{\alpha} \left\{\left(w_\alpha\right)^i_k\left(w_\alpha\right)^j_h
-\left(w_\alpha\right)^j_k\left(w_\alpha\right)^i_h\right\},\label{GMC4}
\end{gather}
\end{subequations}
where $R^{ij}_{kh}=g^{is}R^{j}_{skh}$ are the 
components of the Riemann curvature
tensor of the metric $g$. As observed by 
Ferapontov, if the sum over $\alpha$
goes from 1 to $m$, then these equations
are the Gauss-Mainardi-Codazzi equations 
for an $n$-dimensional submanifold $N$
with flat normal connection embedded in a 
$(n+m)$-dimensional 
pseudo-euclidean space.

It is important to point out that the metric $g$, in general, 
does not uniquely fix the Hamiltonian structure (\ref{ferop});
this arbitrariness is related to the fact that the affinors 
satisfying the Gauss-Peterson-Mainardi-Codazzi equations, and hence the 
corresponding embedding, may not be unique. 

More precisely, let the set of affinors 
$\mathbf{w}\!=\!\{w_\alpha\}$ satisfy equations 
(\ref{gmc}) for a given metric $g$, with 
associated Levi-Civita connection $\nabla$ and 
curvature tensor $R$. 
Let another set of affinors $\mathbf{W}\!=\!\{W_\beta\}$ satisfy the conditions
\begin{gather*}
\left[W_{\beta},w_\alpha\right]=0,\quad
\left[W_{\beta},W_{\beta'}\right]=0,\\
g_{ik}(W_{\beta})^k_j=g_{jk}(W_{\beta})^k_i,\\
\nabla_k(W_{\beta})^i_j=\nabla_j(W_{\beta})^i_k,\\
\quad\qquad\qquad\sum_{\beta}\epsilon_{\beta} 
\left\{\left(W_{\beta}\right)^i_k\left(W_{\beta}\right)^j_h
-\left(W_{\beta}\right)^j_k\left(W_{\beta}\right)^i_h\right\}=0, 
\quad \epsilon_\beta=\pm 1.\label{genzercurv}
\end{gather*}  
It then follows trivially that the union 
$\mathbf{w}\cup \mathbf{W}$ also satisfies equations 
(\ref{gmc}) with the same metric $g$. 
This means that the expression
\begin{align*}
P^{ij}=&\,g^{ij}\frac{d}{dx}-g^{is}\Gamma^j_{sk}u^k_x
+\sum_{\alpha}\varepsilon_{\alpha}\left(w_\alpha\right)^i_ku^k_x
\left(\frac{d}{dx}\right)^{\!-1}\!\!\!\left(w_\alpha\right)^j_hu^h_x\\
&+\sum_{\beta}\epsilon_{\beta}
\left(W_{\beta}\right)^i_ku^k_x
\left(\frac{d}{dx}\right)^{\!-1}\!\!\!\left(W_{\beta}\right)^j_h u^h_x
\end{align*}
is a Poisson bivector; further, it is compatible
with (\ref{ferop}) as one can easily check 
by rescaling $W\to\lambda W$.
In this way one obtains a pencil 
$P_2-\lambda P_1$ of Hamiltonian structures,
where $P_2$ is the Hamiltonian structure of 
Ferapontov type (\ref{ferop})
and $P_1$ is a {\em{purely nonlocal}} Hamiltonian structure of the form 
\beq\label{PNLHS}  
P_1^{ij}=\sum_{\beta}\epsilon_{\beta}\left(W_{\beta}\right)^i_ku^k_x
\left(\frac{d}{dx}\right)^{\!-1}\!\!\!\left(W_{\beta}\right)^j_hu^h_x.
\eeq
Summarizing, the study of the arbitrariness of the nonlocal tail in 
Hamiltonian operators of Ferapontov type (\ref{ferop}) leads us to consider
nonlocal operators of the form (\ref{PNLHS}). 
Apart from purely nonlocal structures associated with \emph{flat} 
metrics considered by Mokhov in \cite{mo} and a few isolated examples \cite{so84,Ch},
such operators have not been 
considered much in the literature. A more systematic study of some 
such structures, a subclass of Mokhov's, was given recently 
in \cite{Pav}, where it was shown that some such purely nonlocal 
Poisson operators can appear as inverses of local symplectic operators.
The aim of this paper is to study Poisson operators of the form (\ref{PNLHS})
in greater generality, to find the conditions they must satisfy, 
and to construct some classes of examples. In the case of a semihamiltonian
hierarchy, there is a remarkable relationship between the symmetries 
$W_{\alpha}$ appearing in the operator, and the metrics naturally associated
with the hierarchy - these are expanded as quadratic forms in the $W_\alpha$.
In particular we find such operators associated with reductions of the 
Benney equations, and with semisimple Frobenius manifolds admitting a superpotential.


\section{Purely nonlocal Hamiltonian 
formalism of hydrodynamic type}\label{purfor} 
Let us consider purely nonlocal operators of the 
form (\ref{PNLHS}); the aim of this section 
is to determine necessary and sufficient 
conditions for (\ref{PNLHS}) to be a Poisson 
operator, namely to satisfy the skew symmetry 
condition and the Jacobi identity. For this 
purpose it is more convenient to consider, 
instead of the differential operator (\ref{PNLHS}), 
its associated bracket
\begin{align}\label{bracket}
\left\{F,G\right\}=&\int\frac{\delta F}{\delta u^i}\,P^{ij}\,\frac{\delta G}{\delta u^j}\,dx\notag\\
=&\iint \frac{\delta F}{\delta u^i(x)}\,\Pi^{ij}(x,y)\,\frac{\delta G}{\delta u^j(y)}\,dy\,dx,
\end{align}
where we have introduced 
\beq\label{candidate}
\Pi^{ij}(x,y)=\sum_\alpha\epsilon_\alpha \,W_\alpha(x)^i_s\,u^s_x\,\,\nu(x-y)\,\,W_\alpha(y)^j_l \,u^l_y,
\eeq
and 
\beq
\nu(x-y)=\f{1}{2}\mbox{sgn}(x-y).
\eeq
The functionals $F$ and $G$ appearing in the 
bracket are of local type, 
not depending on the $x$-derivatives of $u$. 
The skew symmetry of this 
bracket is trivially satisfied, so
we need to find the conditions on the 
$W_\alpha$ such that the Jacobi identity
$$\left\{\left\{G,H\right\}, F\,\right\}+\left\{\left\{F,G\right\},H\,\right\}
+\left\{\left\{H,F\right\}, G\,\right\}=0$$
holds for every $F,$ $G,$ $H$. 

\begin{prop}\label{proppnb}
Suppose that the affinors $W_{\alpha}$ are not degenerate 
and that they 
have a simple spectrum, then the bracket (\ref{bracket}) with 
(\ref{candidate}) satisfies the Jacobi identity 
if and only if the 
following conditions are satisfied:
\begin{gather}
(W_\beta)^m_q\de_m(W_\alpha)^k_l+(W_\beta)^m_l\de_m(W_\alpha)^k_q+(W_\alpha)^k_m\de_l(W_\beta)^m_q+(W_\alpha)^k_m\de_q(W_\beta)^m_l=
\notag\\
\notag\\
=(W_\alpha)^m_q\de_m(W_\beta)^k_l+(W_\alpha)^m_l\de_m(W_\beta)^k_q+(W_\beta)^k_m\de_l(W_\alpha)^m_q+(W_\beta)^k_m\de_q(W_\alpha)^m_l
\label{symmetry}\\
\notag\\
\left[W_\alpha,W_\beta\right]=0, \qquad \forall\, \alpha, \beta,\label{commutativity}\\
\notag\\
\sa\Big((W_\alpha)^i_k(W_\alpha)^j_h-(W_\alpha)^i_h(W_\alpha)^j_k\Big)=0.\label{zerocurv}
\end{gather}
\end{prop}
\begin{rmk}
The first two conditions say that the 
flows $\,\,u^i_{t_{\alpha}}\!=\!(W_\alpha)^i_j\, u^j_x$ commute.
\end{rmk}

\vskip 5mm
\noindent
\pf 
As noticed in \cite{dn89}, in order to prove the 
Jacobi identity we can restrict our attention to 
{\em linear} functionals of the type
$$F=\int f_i(x)\,u^i\,dx,\quad G=\int g_i(x)\,u^i\,dx \quad H=\int h_i(x)\,u^i\,dx.$$
We have
\begin{align}\label{doublebracket}
\left\{\left\{G,H\right\}, F\,\right\}&=\iint \frac{\delta\{G,H\}}{\delta u^m(t)}\,\Pi^{mi}(t,x)\,\frac{\delta F}{\delta u^i(x)}\,dx\,dt\notag\\
&\notag\\
&=\iint \frac{\delta\{G,H\}}{\delta u^m(t)}\,\Pi^{mi}(t,x)\,f_i(x)\,dx\,dt,
\end{align}
where
\begin{align*}
\frac{\delta\{G,H\}}{\delta u^m(t)}&=\frac{\delta}{\delta u^m(t)}\iint \frac{\delta G}{\delta u^j(y)}\,\Pi^{jk}(y,z)\,\frac{\delta H}{\delta u^k(z)}\,dy\,dz\\
&\\
&=\iint g_j(y)\,\frac{\delta\,\Pi^{jk}(y,z)}{\delta u^m(t)}\,h_k(z)\,dy\,dz.
\end{align*}
Hence, (\ref{doublebracket}) can be reduced to 
$$\left\{\left\{G,H\right\}, F\,\right\}=\iiint f_i(x)\,g_j(y)\,h_k(z)\,S^{ijk}(x,y,z)\,dx\,dy\,dz,$$
where we have  introduced the quantity
$$S^{jki}(y,z,x)=\int\frac{\delta\,\Pi^{jk}(y,z)}{\delta u^m(t)}\,\Pi^{mi}(t,x)\,dt.$$
In this way, the Jacobi identity reads
$$\iiint f_i(x)\,g_j(y)\,h_k(z)\,\Big(S^{jki}(y,z,x)+S^{kij}(z,x,x)+S^{ijk}(x,y,z)\Big)\,dx\,dy\,dz=0,$$
This has to be satisfied for every function $f_i$, $g_j$, $h_k$, so that we must require 
\beq\label{jacs}
S^{jki}(y,z,x)+S^{kij}(z,x,x)+S^{ijk}(x,y,z)=0.
\eeq

Let us consider the quantities $S^{jki}(y,z,x)$ more explicitly. 
We have
\begin{align*}
\frac{\delta\,\Pi^{jk}(y,z)}{\delta u^m(t)}&=\sum_\alpha\epsilon_\alpha\left\{\left[\frac{\de W_\alpha(y)^j_p}{\de u^m(t)}\,\delta(y-t)\,u^p_y+
W_\alpha(y)^j_m\,\delta'(y-t)\right]W_\alpha(z)^k_q \,u^q_z\right.\\
&\\
&+\left.W_\alpha(y)^j_p\,u^p_y\left[\frac{\de W_\alpha(z)^k_q}{\de u^m(t)}\,\delta(z-t)\,u^q_z+
W_\alpha(z)^k_m\,\delta'(z-t)\right]\right\}\,\nu(y-z), 
\end{align*}
and so
\begin{align*}
S^{jki}(y,z,x)&=\left(\sum_\alpha\epsilon_\alpha\frac{\de W_\alpha(y)^j_p}{\de u^m(y)}\,W_\alpha(z)^k_q\right)\,\Pi^{mi}(y,x)\,u^p_y \,u^q_z\,\nu(y-z)\\
&\\
&+\left(\sum_\alpha\epsilon_\alpha W_\alpha(y)^j_m\,W_\alpha(z)^k_q\right)\frac{\de\Pi^{mi}(y,x)}{\de y} \,u^q_z\,\nu(y-z)\\
&\\
&+\left(\sum_\alpha\epsilon_\alpha W_\alpha(y)^j_p\,\frac{\de W_\alpha(z)^k_q}{\de u^m(z)}\right)\,\Pi^{mi}(z,x)\,u^p_y\,u^q_z\,\nu(y-z)\\
&\\
&+\left(\sum_\alpha\epsilon_\alpha W_\alpha(y)^j_p\,W_\alpha(z)^k_m\right)\,\frac{\de\Pi^{mi}(z,x)}{\de z}\,u^p_y\,\nu(y-z).
\end{align*}
Now we evaluate:
\begin{align*}
\frac{\de\Pi^{mi}(y,x)}{\de y}=&\left(\sa\frac{\de W_\alpha(y)^m_p}{\de u^l(y)}\,W_\alpha(x)^i_s\right)
u^s_x\,u^p_y\,u^l_y\,\nu(y-x)\\
&\\
&+\left(\sa W_\alpha(y)^m_p\,W_\alpha(x)^i_s\right)u^s_x\,u^p_{yy}\,\nu(y-x)\\
&\\
&+\left(\sa W_\alpha(y)^m_p\,W_\alpha(x)^i_s\right)u^s_x\,u^p_y\,\delta(y-x),
\end{align*}
so that we can write
\begin{align*}
S^{jki}(y,z,x)&=A^{kji}_{qpls}(z,y,x)\,u^s_x\,u^p_y\,u^l_y\,u^q_z\,\nu(y-x)\,\nu(y-z)\\
&\\
&+A^{jki}_{pqls}(y,z,x)\,u^s_x\,u^p_y\,u^q_z\,u^l_z\,\nu(z-x)\,\nu(y-z)\\
&\\
&+B^{kji}_{qps}(z,y,x)\,u^s_x\,u^p_{yy}\,u^q_z\,\nu(y-x)\,\nu(y-z)\\
&\\
&+B^{jki}_{pqs}(y,z,x)\,u^s_x\,u^p_y\,u^q_{zz}\,\nu(z-x)\,\nu(y-z)\\
&\\
&+B^{kji}_{qps}(z,y,x)\,u^s_x\,u^p_y\,u^q_z\,\delta(y-x)\,\nu(y-z)\\
&\\
&+B^{jki}_{pqs}(y,z,x)\,u^s_x\,u^p_y\,u^q_z\,\delta(z-x)\,\nu(y-z),
\end{align*}
Here we have introduced the notation:
\begin{align}\label{defA}
A^{kji}_{qpls}(z,y,x)&=\left(\sa W_\alpha(z)^k_q\de_m W_\alpha(y)^j_p\right)\left(\sa W_\alpha(y)^m_l\,W_\alpha(x)^i_s\right)+\notag\\
&\notag\\
&+\left(\sa W_\alpha(z)^k_q\, W_\alpha(y)^j_m\right)\left(\sa\de_{\,l}\, W_\alpha(y)^m_p\,W_\alpha(x)^i_s\right),
\end{align}
\begin{align}\label{defB}
B^{kji}_{qps}(z,y,x)=\left(\sa W_\alpha(z)^k_q\, W_\alpha(y)^j_m\right)\left(\sa W_\alpha(y)^m_p\,W_\alpha(x)^i_s\right).
\end{align}
Cyclically permuting with respect to $i,j,k$ and $x,y,z$, 
and then rearranging the terms, it is possible to 
rewrite condition (\ref{jacs}) in the form
\begin{align*}
&\left[A^{ijk}_{splq}(x,y,z)-A^{kji}_{qpls}(z,y,x)\right]\,u^s_x\,u^p_y\,u^l_y\,u^q_z\,\nu(x-y)\,\nu(y-z)\\
&\\
+&\left[B^{ijk}_{spq}(x,y,z)-B^{kji}_{qps}(z,y,x)\right]\,u^s_x\,u^p_{yy}\,u^q_z\,\nu(x-y)\,\nu(y-z)\\
&\\
+&\left[B^{ijk}_{spq}(x,y,z)-B^{kji}_{qps}(z,y,x)\right]\,u^s_x\,u^p_y\,u^q_z\,\delta(x-y)\,\nu(y-z)\\
&\\
+&\left(\text{cyclic permutations}\right)=0,
\end{align*}
and this reduces to the following conditions 
(for if these are satisfied, then all the others vanish identically):
\begin{gather}
A^{ijk}_{splq}(x,y,z)+A^{ijk}_{slpq}(x,y,z)=A^{kji}_{qpls}(z,y,x)+A^{kji}_{lpqs}(z,y,x),\\
\notag\\
B^{ijk}_{spq}(x,y,z)=B^{kji}_{qps}(z,y,x)\\
\notag\\
B^{ijk}_{spq}(x,x,z)+B^{ijk}_{psq}(x,x,z)=B^{kji}_{qps}(z,x,x)+B^{kji}_{qsp}(z,x,x)
\end{gather}
that follow immediately from (\ref{symmetry}), (\ref{commutativity}), (\ref{zerocurv}).
The converse is also true if the affinors $W_\alpha$ 
are not degenerate, and have a simple spectrum.  

\epf

\section{Semi-Hamiltonian systems}

Let us consider now the important class of diagonalizable, 
semi-Hamiltonian systems of hydrodynamic type. 
These systems were introduced by Tsarev in \cite{ts91}, 
and correspond to the class of systems which are integrable 
by the {\em generalized hodograph method}. 
More precisely, a diagonal system of hydrodynamic type 
\beq\label{diag}
u^i_t=v^i(u)u^i_x,
\eeq
\noindent
is called \emph{semi-Hamiltonian} \cite{ts91}  if the coefficients 
$v^i(u)$ satisfy the system of equations
\begin{equation}\label{sh}
\partial_j\left(\frac{\partial_k v^i}{v^i-v^k}\right)=
\partial_k\left(\frac{\partial_j v^i}{v^i-v^j}\right)\hspace{1
cm}\forall\, i\ne j\ne k\ne i,
\end{equation}
where $\de_i=\frac{\partial}{\partial \lm^i}$. 
The $v^i$ are usually called \emph{characteristic velocities}. 
Equations (\ref{sh}) are the integrability conditions for 
three different systems: 
the first, given by
\begin{equation}
\label{SYM} \frac{\de_j w^i}{w^i-w^j}=\frac{\de_j v^i}{v^i-v^j},
\end{equation}
which provides  the characteristic velocities $w^i(u)$ of the symmetries
of (\ref{hts}):
\begin{equation*}
u^i_{\tau}=w^i(u)u^i_x\hspace{1 cm}i=1,...,n;
\end{equation*}
the second is a system whose solutions are the conserved densities $H(u)$
 of (\ref{hts}):
\begin{equation}\label{conl}
\d_i\d_j H-\Gamma^i_{ij}\d_i H-\Gamma^j_{ji}\d_j H=0,
\qquad\Gamma^i_{ij}=\f{\de_j v^i}{v^j-v^i},
\end{equation}
and the third is
\begin{equation}\label{meq} 
\de_j\ln{\sqrt{g_{ii}}}=\frac{\de_j v^i}{v^j-v^i},\qquad i\neq j,
\end{equation}
which relates the characteristic velocities of the system 
to a class of diagonal metrics $g_{ii}(u)$. 
In \cite{fe91} Ferapontov noticed that these metrics represent 
all possible candidates for the construction of Hamiltonian 
operators for the system, whether of local type (\ref{PBHT}), 
or nonlocal (\ref{ferop}). 

Now let us consider a purely nonlocal Hamiltonian formalism. Let
\beq\label{sh2}
u^i_t=v^i u^i_x,
\eeq
be a semi-Hamiltonian system and let
$$u^i_{t_{\alpha}}=W^i_{\alpha}u^i_x,$$
be a set of symmetries satisfying condition \eqref{zerocurv}; if the affinors
are diagonal, this takes the form:
\begin{gather}
\sa W^i_\alpha W^j_\alpha=0 \qquad i\neq j. \label{orth}
\end{gather}
Then, according to the results of Section \ref{purfor}, the operator
\beq\label{pnhs}
P=\sa W^i_\alpha\, u^i_x\left(\frac{d}{dx}\right)^{-1}   W^j_\alpha\,u^j_x
\eeq
defines a Hamiltonian structure. 
Moreover, the flows generated by the Hamiltonian densities 
which solve the linear system \eqref{conl} 
are symmetries of \eqref{sh2}. 

More precisely, we consider 
$$u^i_\tau=w^iu^i_x, \qquad w^i:=P^{ij}\de_jH=\sa W^i_\alpha K^\alpha,$$
where the functions $K^\alpha$ are the fluxes of 
conservation laws given by
\beq\label{currents}
\de_{t_\alpha}H=\de_x K^\alpha.
\eeq
For $i\neq j$ we get:
$$\de_j w^i=\de_j\left(\sa W^i_\alpha K^\alpha\right)=\sa \de_j W^i_\alpha K^\alpha+\sa W^i_\alpha\de_j K^\alpha,$$
so by using equations \eqref{SYM} and \eqref{currents} we obtain
\begin{align*}
\de_j w^i&=\frac{\de_j v^i}{v^j-v^i}\sa\left(W^j_\alpha-W^i_\alpha\right)K^\alpha+\sa W^i_\alpha W^j_\alpha \de_i H\\
&= \frac{\de_j v^i}{v^j-v^i}\left(w^j-w^i\right).
\end{align*} 
Hence the Hamiltonian flows generated by conserved densities 
indeed belong to the semi-Hamiltonian hierarchy containing \eqref{sh2}. 
The converse problem, namely whether an arbitrary flow
\beq\label{sh2s}
u^i_{\tau}=X^i=w^i u^i_x
\eeq
commuting with (\ref{sh2}) is Hamiltonian 
with respect to the purely nonlocal Poisson structure \eqref{pnhs}, 
turns out to be much more difficult to solve.  
However, we can say that the the Hamiltonian structure 
\eqref{pnhs} is conserved along any flow \eqref{sh2s} 
of the hierarchy: 
\begin{prop}
Let 
\beq\label{pbivpf}
\Pi^{ij}(x,y)=\sum_{\alpha}\,W^i_{\alpha}(x)\,u^i_x\,\nu(x-y)\,W^j_{\alpha}(y)\,u^j_y
\eeq
be a purely nonlocal Poisson bivector, and suppose that the 
commuting flows
$$u^i_{t_\alpha}=W^i_{\alpha}(u^1(x),\dots,u^n(x))u^i_x,$$
belong to a semi-Hamiltonian hierarchy. 
If
\beq\label{vecfielpf}
u^i_t=X^i(x)=w^i(x)u^i_x,
\eeq
is an arbitrary flow of this hierarchy, then
$${\rm Lie}_X \Pi=0.$$
\end{prop}

\noindent
\emph{Proof}. For the bivector \eqref{pbivpf} and the 
vector field \eqref{vecfielpf} it is not difficult to show \cite{DZ} 
that the expression ${\rm Lie}_X \Pi$ takes the form:
\begin{eqnarray*}
\left[{\rm Lie}_X \Pi\right]^{ij}&=&
X^k(x)\,\f{\d \Pi^{ij}(x,y)}{\d u^k(x)}+\d_x X^k(x)\,\f{\d \Pi^{ij}(x,y)}{\d u^k_x}\\
&&X^k(y)\,\f{\d \Pi^{ij}(x,y)}{\d u^k(y)}+\d_y X^k(y)\,\f{\d \Pi^{ij}(x,y)}{\d u^k_y}\\
&&-\f{\d X^i(x)}{\d u^k(x)}\,\Pi^{kj}(x,y)-\f{\d X^i(x)}{\d u^k_x}\,\d_x \Pi^{kj}(x,y)\\
&&-\f{\d X^j(y)}{\d u^k(y)}\,\Pi^{ik}(x,y)-\f{\d X^j(y)}{\d u^k_y}\,\d_y \Pi^{ik}(x,y).
\end{eqnarray*}
Rearranging and simplifying, we obtain
\begin{eqnarray*}
\left[{\rm Lie}_X \Pi\right]^{ij}&=&
\sum_{\alpha}\,w^k(x)\,u^k_x\,\f{\d W^i_{\alpha}(x)}{\d u^k(x)}\,u^i_x\,\nu(x-y)\,W^j_{\alpha}(y)\,u^j_y\\
&&+\sum_{\alpha}\left[w^k(x)\,u^k_{xx}+\f{\d w^k(x)}{\d u^l(x)}\,u^k_x\,u^l_x\right]\,\delta^i_k\,W^i_{\alpha}(x)\,\nu(x-y)\,W^j_{\alpha}(y)\,u^j_y\\
&&+\sum_{\alpha}\,w^k(y)\,u^k_y\,W^i_{\alpha}(x)\,u^i_x\,\nu(x-y)\,\f{\d W^j_{\alpha}(y)}{\d u^k(y)}\,u^j_y\\
&&+\sum_{\alpha}\left[w^k(y)\,u^k_{yy}+\f{\d w^k(y)}{\d u^l(y)}\,u^k_y\, u^l_y\right]\,\delta^j_k\,W^i_{\alpha}(x)\,u^i_x\,\nu(x-y)\,W^j_{\alpha}(y)\\
&&-\sum_{\alpha}\,\f{\d w^i(x)}{\d u^k(x)}\,u^i_x\,W^k_{\alpha}(x)\,u^k_x\,\nu(x-y)\,W^j_{\alpha}(y)\,u^j_y\\
&&-\sum_{\alpha}\,\delta^i_k\,w^i(x)\left[W^k_{\alpha}(x)\,u^k_{xx}+\f{\d W^k_{\alpha}(x)}{\d u^l(x)}\,u^k_x\,u^l_x\right]\,\nu(x-y)\,W^j_{\alpha}(y)\,u^j_y\\
&&-\sum_{\alpha}\,\delta^i_k\,w^i(x)\,W^k_{\alpha}(x)\,u^k_x\,\delta(x-y)\,W^j_{\alpha}(y)\,u^j_y\\
&&-\sum_{\alpha}\,\f{\d w^j(y)}{\d u^k(y)}\,u^j_y\,W^i_{\alpha}(x)\,u^i_x
\,\nu(x-y)\,W^k_{\alpha}(y)\,u^k_y\\
&&-\sum_{\alpha}\,\delta^j_k\,w^j(y)\,W^i_{\alpha}(x)\,u^i_x\,\nu(x-y)\,\left[W^k_{\alpha}(y)\,u^i_{yy}+\f{\d W^k_{\alpha}(y)}{\d u^l(y)}\,u^k_y \,u^l_y\right]\\
&&+\sum_{\alpha}\,\delta^j_k\,w^j(y)\,W^i_{\alpha}(x)\,u^i_x\,\delta(x-y)\,W^k_{\alpha}(y)\,u^k_y.
\end{eqnarray*}
The terms containing the second derivatives vanish identically. 
Collecting the remaining terms and using the properties of the 
delta function we obtain
 \begin{eqnarray*}
&&\left[{\rm Lie}_X \Pi\right]^{ij}=
\left[w^j(x)-w^i(x)\right]
\left(\sum_{\alpha}W^i_{\alpha}(x)W^j_{\alpha}(x)\right)
u^i_x\, u^j_x\,\delta(x-y)\\
 &&+\sum_{\alpha}\left[\f{\d W^i_{\alpha}(x)}{\d u^k(x)}\left(w^k(x)-w^i(x)\right)-\f{\d w^i(x)}{\d u^k(x)}\left(W^k_{\alpha}(x)-
 W^i_{\alpha}(x)\right)\right]u^k_x u^i_x\nu(x-y)W^j_{\alpha}(y)u^j_y\\
&&+\sum_{\alpha}\left[\f{\d W^j_{\alpha}(y)}{\d u^k(y)}\left(w^k(y)-w^j(y)\right)-
 \f{\d w^j(y)}{\d u^k(y)}\left(W^k_{\alpha}(y)-
 W^j_{\alpha}(y)\right)\right]W^i_{\alpha}(x)u^i_x \nu(x-y)u^j_y u^k_y, 
 \end{eqnarray*}
which is identically zero, because:
$$\sum_{\alpha}W^i_{\alpha}(x)W^j_{\alpha}(x)=0, \qquad i\neq j,$$
and
$$\f{\d_k w^i}{w^k-w^i}=\f{\d_k W^i_{\alpha}}{W^k_{\alpha}-W^i_{\alpha}}, \qquad k\neq i.$$
Thus, indeed,
$${\rm Lie}_X \Pi=0.$$
\begin{flushright}
$\Box$
\end{flushright}

\section{Quadratic expansion of the metric}

Remarkably, in the case of semi-Hamiltonian systems, 
the existence of purely nonlocal Poisson structures 
is related to a quadratic expansion 
\begin{equation}\label{expconmet}
g^{ii}\delta^{ij}=\sa W^i_\alpha W^j_\alpha,
\end{equation}
of the {\em contravariant} 
components of a metric $g$,  whose covariant components 
satisfy \eqref{meq}, namely:
$$\de_j\ln{\sqrt{g^{ii}}}=
-\frac{\de_j v^i}{v^j-v^i},\qquad i\neq j.$$

For $i\neq j$, the former identity follows from \eqref{orth}, 
while for the diagonal components we have the following
\begin{prop}\label{propmet}
Let a diagonal system (\ref{diag}) be semi-Hamiltonian, 
and suppose we have a set of symmetries $W_\alpha$ satisfying 
condition (\ref{orth}) for certain $\epsilon_\alpha=\pm 1$. 
Then, the set of functions
$$Q^{\,i}:=\sa \left(W_\alpha^i\right)^2,$$
satisfies the system
$$\de_j\ln{\sqrt{Q^{\,i}}}=-\frac{\de_j v^i}{v^j-v^i}, 
\qquad i\neq j.$$ \end{prop}
\noindent
\emph{Proof}.
For $i\neq j,$ we have 
\begin{align*}
\de_j Q^{\,i}=&\,\,\de_j\left(\sum_{\alpha}\epsilon_{\alpha}(W^i_{\alpha})^2\right)
=2\,\sum_{\alpha}\epsilon_{\alpha}W^i_\alpha\,\de_jW^i_\alpha\\
=&\,\,2\,\sum_{\alpha}\epsilon_{\alpha}W^i_\alpha\left(W^j_\alpha-W^i_\alpha\right)\frac{\de_jv^i}{v^j-v^i}
=-2\,Q^{\,i}\,\frac{\de_jv^i}{v^j-v^i}, \qquad i\neq j.
\end{align*}
\begin{flushright}
$\Box$
\end{flushright}

\begin{rmk}\label{egorov}
If $g_{ii}$ is a metric of Egorov type, that is 
$$g_{ii}=\d_i H$$ 
for a suitable function $H$, then it is known (see e.g. \cite{PvTs}) that the characteristic velocities $W^i_{\alpha}$ can be written as
\begin{equation}\label{reps}
W^i_{\alpha}=-\f{\d_i K_{\alpha}}{\d_i H}=-\f{\d_i K_{\alpha}}{g_{ii}},
\end{equation}
where the $K_{\alpha}$ are densities of conservation laws. 
In this case, equation \eqref{expconmet} can be written as
$$g^{ii}\delta^{ij}=\frac{1}{g_{ii}\,g_{jj}}\sa \d_i K_{\alpha} \d_j K_{\alpha},$$
that is
$$g_{ii}\,\delta_{ij}=\sa \d_i K_{\alpha} \d_j K_{\alpha}.$$
This suggests that, in the case of Egorov metrics, the existence of
 a quadratic expansion for a solution of the linear system (\ref{meq})
 is related to the existence of an embedding of our $n$ dimensional 
 manifold $N$ in a pseudo-euclidean space with coordinates 
 $K_{\alpha}$, in 
which the metric $g$ plays the role of the first fundamental form.  
\end{rmk}

We have proved that any purely nonlocal 
Hamiltonian structure constructed for a semi-Hamiltonian 
system is necessarily related with one of the metrics which solves 
\eqref{meq}. This relation has been obtained with a direct 
calculation using the diagonal coordinates frame, 
in which both the symmetries and the metric are diagonal.

In order to give a coordinate-free formulation of this, it is
convenient to interpret the characteristic velocities of the symmetries 
entering in the quadratic expansion of the metric as a vector fields on 
our $n$-dimensional manifold $N$. 
Such a change of point of view leads us naturally to introduce an  
algebraic structure on the tangent bundle $TN$ of our $n$-dimensional
manifold $N$ - each fibre $T_uN$ has the structure of an associative 
semisimple multiplicative
algebra, and the bundle  admits a holonomic basis of idempotents. 
This means that first, there exists a basis $(Z_1,\dots,Z_n)$
of idempotents:
$$
Z_i(u) \cdot Z_j(u)=\delta_{ij} Z_j(u).$$
Second, if  this basis commutes, (is holonomic), then there exists a 
set of coordinates, called {\em{canonical coordinates}}, 
$(u^1,\dots,u^n)$ such that
$$ Z_i=\frac{\partial}{\partial u^i}.$$

An invariant form of the condition (3.1) can now be easily obtained by noting
the following.
\begin{lemma}
Any diagonalizable system of hydrodynamic type 
\beq\label{shtc}
u^i_t=v^i_j(u)u^j_x, 
\eeq
can be written in the form
\beq\label{shtcW}
v^i_j(u)=c^i_{\,jk}(u)X^k(u),
\eeq
where the $X^k$ are now the components of a vector field and the 
$c^i_{\,jk}$ are the ($u$-dependent) structure ``constants'' of a
associative semisimple algebra admitting a holonomic basis of idempotents.
\end{lemma} 

\noindent
\emph{Proof}. 
\noindent
Indeed system (\ref{shtcW}) becomes diagonal in canonical 
coordinates -  and in such coordinates, the structure 
constants are simply
\beq\label{sccc}
c^i_{\,jk}=\delta^i_j\delta^i_k.
\eeq
These $c^i_{\,jk}$ are evidently the structure constants of an 
associative algebra.
Conversely, given a diagonal system of hydrodynamic type, we 
can define the structure constants by identifying the canonical 
coordinates with the given  Riemann invariants by means of 
(\ref{sccc}). This identification will
clearly depend on the choice of the Riemann invariants.
\begin{flushright}
$\Box$
\end{flushright}

\begin{thm}
Let
\beq\label{sac}
u^i_{t_{\alpha}}=(W_{\alpha})^i_j(u)u^j_x=c^i_{jk}(u)X_{\alpha}^k(u)u^j_x
\eeq
be $n$ commuting diagonal systems of hydrodynamic type defined by the 
structure  constants of an associative semisimple algebra, admitting a 
holonomic basis of idempotents and by $n$ vector fields $X_{\alpha}$ ($\alpha=1,\dots,m$).

Suppose that the metric
\beq\label{c1}    
g^{ij}=(\sum_{\alpha=1}^n\epsilon_{\alpha}X_{\alpha}\otimes X_{\alpha})^{ij}.
\eeq
is nondegenerate and satisfies the following condition 
\beq\label{c2}
g^{kl}c^i_{\,lm}=g^{il}c^k_{\,lm}.
\eeq
Then:

1. Denoting by $\nabla$ the Levi-Civita connection associated with $g$, 
we have 
\beq\label{c3}
g_{ik}(W_{\alpha})^k_j=
g_{jk}(W_{\alpha})^k_i,\hspace{1 cm}\nabla_k (W_{\alpha})^i_j=\nabla_j (W_{\alpha})^i_k
\eeq

2. The affinors $(W_{\alpha})^i_j$ satisfy the conditions (1.4,1.5,1.6) 
and therefore the operator  
$$\sum_{\alpha=1}^M\epsilon_{\alpha}\left(W_{\alpha}\right)^i_ku^k_x
\left(\frac{d}{dx}\right)^{\!-1}\!\!\!\left(W_{\alpha}\right)^j_hu^h_x$$
is a purely nonlocal Hamiltonian operator.
\end{thm}

\noindent
\emph{Proof}. 

\noindent
1. Condition (\ref{c2}) implies that the metric (\ref{c1}) is diagonal in
 canonical coordinates. Moreover it implies the first of conditions (\ref{c3}).
  In order to prove the second of conditions (\ref{c3}) we observe
   that, in canonical coordinates, it reads
 $$\de_j\ln{\sqrt{g_{ii}}}=
\frac{\de_j W_{\alpha}^i}{W^j_{\alpha}-W^i_{\alpha}},\qquad i\neq j.$$
Taking into account that, in canonical coordinates, the vector fields  
$W_{\alpha}$ ($\alpha=1,\dots,m$) coincide with the characteristic 
velocities  of the systems (\ref{sac}) and satisfy the condition (\ref{orth}),
 we obtain the result by computations of the Proposition \ref{propmet}. 

\noindent 
2. Conditions (\ref{symmetry},\ref{commutativity}) follow from the commutativity of the flows (\ref{sac}). 
Condition (\ref{zerocurv}) follows immediately from (\ref{c1},\ref{c2}). Indeed
\begin{eqnarray*}
&&\sum_{\alpha}\epsilon_{\alpha}\left((W_{\alpha})^i_{k}(W_{\alpha})^j_{h}-(W_{\alpha})^i_{h}(W_{\alpha})^j_{k}\right)=
\sum_{\alpha}\epsilon_{\alpha}(c^i_{\,kl}c^j_{\,hm}-c^i_{\,hl}c^j_{\,km})(X_{\alpha})^l(X_{\alpha})^m=\\
&&=(c^i_{\,kl}c^j_{\,hm}-c^i_{\,hl}c^j_{\,km})g^{lm}.
\end{eqnarray*}
Using (\ref{c2}) we get
$$(c^i_{\,kl}c^j_{\,hm}-c^i_{\,hl}c^j_{\,km})g^{lm}
=g^{js}(c^i_{\,kl}c^l_{\,hs}-c^i_{\,hl}c^l_{\,ks})$$
which vanishes due to associativity. 

\begin{flushright}
$\Box$
\end{flushright}
\begin{rmk}
We should point out that the second part of the theorem only uses the associativity
property; the assumption of semisimplicity is only needed for the first part,
which uses canonical coordinates.
\end{rmk}

\section{Reductions of the Benney system}
We recall the basic facts about the Benney chain and its reductions (for
details see for example \cite{glr2008} and references therein). 
The Benney chain is the following infinite system of quasilinear PDEs:
\beq\label{bmc}
A^k_t=A^{k+1}_x+kA^{k-1}A^0_x,\qquad k=0,1,2,\dots,
\eeq
in the infinitely many variables $A^k(x,t)$, 
which are usually called moments.
Introducing the formal series
$$\lm=p+\sum_{k=0}^{\infty}\frac{A^k}{p^{k+1}},$$
we can encode the whole system in the single equation
\beq\label{dKP} 
\lambda_t=p\lambda_x-A^0_x\lambda_p,
\eeq
which is the second flow of the dispersionless $KP$ hierarchy; by 
considering the inverse of $\lm$ with respect to $p$, we obtain the 
series 
\beq\label{pseries}
p=\lambda-\sum_{k=0}^{\infty}\frac{H_k}{\lambda^{k+1}},
\eeq
whose coefficients are conserved densities of the Benney chain, each of 
them polynomial in the moments.
\begin{rmk}
In many important examples, and in particular for all the reductions 
defined below, the series $\lm$ can be thought as the asymptotic 
expansion at infinity of an analytic function $\lm(p,x,t)$. In this 
case, the generating function (\ref{pseries}) is obtained by 
inverting the function $\lm$ with respect to $p$, and then expanding 
asymptotically around infinity.
\end{rmk}

A \emph{reduction} of the Benney chain is a suitable restriction of the 
system (\ref{bmc}) to the case when all the moments $A^k$ can be 
expressed in terms of finitely many variables; as proved in \cite{GT}, 
all reductions of the Benney chain are diagonalizable, 
that is they can be written in the form 
 \beq\label{rbs}
\lambda^i_t=v^i(\lambda)\lambda^i_x,
\eeq
and they satisfy the semi-Hamiltonian condition (\ref{sh}). As can easily
 be understood, in the case of a reduction the corresponding function 
$\lm$ depends on the variables $x$ and $t$ only through 
$\lm^1,\dots,\lm^n$, that is
$$\lm(p,x,t)=\lm(p,\lm^1(x,t),\dots,\lm^n(x,t)).$$
In this case, and assuming the linear independence of the $\lm^i_x$, 
(\ref{dKP}) is equivalent to the system
\beq\label{loewner}
\frac{\d\lambda}{\d\lambda^j}=
\frac{\frac{\d A^0}{\d \lambda^j}}{p-v^j}\frac{\d\lambda}{\d p},
\qquad\quad j=1,\dots,n,
\eeq
which is a system of $n$ \emph{Loewner equations}, which describe -- 
for instance -- families of conformal maps from the upper complex half plane
 to the upper complex half plane with $n$ arbitrary slits \cite{GT2}. 
The analytic properties of the conformal map solutions of 
(\ref{loewner}) are intimately related to the properties of the 
corresponding reduction. For example, the critical points of $\lambda$ 
are the characteristic velocities $v^i$ and its critical values are 
Riemann invariants:
$$\frac{\de \lambda}{\de p}(v^i)=0, \qquad \lm(v^i)=\lm^i.$$
Moreover, the coefficients of the
expansion at $\lambda=\infty$ of the functions
\beq\label{symmbenn}
W_i(\lambda,\lambda^1,\dots,\lambda^n)=\frac{1}{p(\lambda)-v^i}
=\sum_{n=1}^{\infty}\frac{w^i_{(n)}}{\lambda^n}
\eeq
are characteristic velocities of symmetries. Finally, as proved by the 
present authors in \cite{glr2008}, reductions of the Benney system 
associated with the function 
$\lambda(p,\lambda^1,\dots,\lambda^n)$
are Hamiltonian with respect to the Hamiltonian structures 
\beq\label{Hsb}
P^{ij}=\varphi_i(\lambda^i)\,\lambda^{''}(v^i)\delta^{ij}\frac{d}{dx}+\Gamma^{ij}_k\,\lambda^k_x-\frac{1}{2\pi i}\sum_{l=1}^n\int_{C_l}w^i(\lambda)\lambda^i_x\left(\frac{d}{d
x}\right)^{-1}w^j(\lambda)\lambda^j_x\,\varphi_l(\lambda)\,d\lambda
\eeq
where the contour $C_i$ is the image of a sufficiently small closed contour
around the point $p=v^i$ in the $p$-plane with respect to the analytic 
continuation of  the conformal map $\lambda(p)$, the functions $\varphi_i$
are arbitrary functions of $\lambda$, and 
$$w^i(\lambda):=-\frac{\frac{\d p}{\d\lm}}{(p(\lm)-v^i)^2}=\frac{\d W_i}{\d\lambda}.$$
As all reductions of the Benney chain are semi-Hamiltonian systems, in 
addition to the Hamiltonian structures (\ref{Hsb}) we can obtain a
family of purely nonlocal Hamiltonian structures if we can expand the 
contravariant components of the metrics
 \beq\label{metben}
 g_{\varphi}^{ii}=\varphi_i(\lambda^i)\lambda^{''}(v^i),
 \eeq
in terms of symmetries of the system.

\begin{thm}
The components of a metric associated with a reduction of the Benney chain admit the following quadratic expansion 
\beq\label{qem}
g_\varphi^{ii}\,\delta^{ij}=\varphi_i(\lambda^i)\lm''(v^i)\delta^{ij}=\frac{1}{2\pi  i}\sum_{k=1}^n\int_{C_k}W_i(\lambda)W_j(\lambda)\varphi_k(\lambda)\,d\lambda,
\eeq
where the $W_i(\lambda)$ are the generating functions of the symmetries (\ref{symmbenn}) and the contours $C_k$ are the same as in (\ref{Hsb}).
\end{thm}

\pf
The proof is a straightforward computation of the integral:
\begin{gather*}
\frac{1}{2\pi i}\sum_{k=1}^n\int_{C_k}W_i(\lambda)W_j(\lambda)\varphi_k(\lambda)\,d\lambda=
\sum_{k=1}^n\underset{\lm=\lm^k}{\rm Res}\left[\frac{
\varphi_k(\lambda)\,d\lm}{(p(\lm)-v^i)(p(\lm)-v^j)}\right]\\
=\sum_{k=1}^n\underset{p=v^k}{\rm Res}\left[\frac{\frac{\de \lm}{\de p}}{(p-v^i)(p-v^j)}\,
\,\varphi_k(\lambda(p))\,\,d p\right]=\varphi_i(\lm^i)\,\lm''(v^i)\delta^{ij},
\end{gather*}
the last step being due to the fact that $p=v^k$ are critical points of 
$\lm$, so that the differential turns out to be regular at all these 
points for $i\neq j$, and also for $i=j$ and $k\neq i$.
\begin{flushright}
$\Box$
\end{flushright}

\begin{rmk} In the Benney case, it is known \cite{glr2008} that the 
metric associated with a reduction are of Egorov type, and more 
precisely of the form 
$$\left(g_\varphi\right)_{ii}=\frac{1}{\varphi_i(\lambda^i)}\,.$$
Moreover, the function $p(\lm)$ satisfies a Loewner system of the form
$$\de_i p=-\frac{\de_i A^0}{p(\lm)-v^i}=-W^i(\lm)\,\de_i A^0,$$
obtained by (\ref{loewner}) by using the implicit function theorem.
Using Remark \ref{egorov} about the Egorov metrics, we conclude that 
the covariant metrics associated with a reduction of the Benney chain 
can be written as
\begin{equation}\label{qexpbr}
\left(g_\varphi\right)_{ii}\delta_{ij}=
\frac{1}{2\pi  i}\sum_{k=1}^n\int_{C_k}\f{\d_i p\d_j p}{\varphi_k(\lambda)}\,d\lambda.
\end{equation}
\end{rmk}

\section{Semi-Hamiltonian systems related to semisimple Frobenius manifolds}

We have seen that these purely nonlocal Hamiltonian structures are 
connected with a geometrical structure where the tangent space of a 
manifold has the structure of an associative multiplicative algebra. 
The most important examples of these are {\em{Frobenius manifolds}}.
A Frobenius manifold \cite{du93,du99} is a manifold $M$ endowed  with a 
commutative, associative multiplicative structure $\,\cdot\,$ on the 
tangent spaces together with a flat metric $\eta$, invariant with 
respect to the product $\,\cdot\,$. 
This means that the third order tensor $c$ defined by
$$c(u,v,w)=(u\cdot v,w)$$
(where $u,v,w$ are arbitrary vector fields and $(\;\,,\;)$ 
is the scalar product defined by $\eta$) is symmetric. 

It is easy to check that this condition, combined with requiring the symmetry
 of the fourth order tensor
$$\nabla_z c(u,v,w)$$
 implies that, in flat coordinates $v^1,\dots,v^n$,
 the structure constants of $\,\cdot\,$
 can be written (locally) as third derivatives of a function $F$, 
  called the \emph{Frobenius potential}:
$$c_{\alpha\beta\gamma}=\eta_{\alpha\delta}c^{\delta}_{\beta\gamma}=
\f{\d^3 F}{\d v^{\alpha}\d v^{\beta} \d v^{\gamma}}.$$

The definition of a Frobenius manifold also involves  two special 
vector fields:  
the first, usually denoted by $e$, is the unit of the 
product $\,\cdot\,$ and can be identified with the vector field 
$\frac{\d}{\d v^1}$; 
the second, called the \emph{Euler vector field},
 encodes the quasi-homegeneity properties of the Frobenius potential $F$:
\beq\label{qop}
Lie_E(F)=(3-d)F,
\eeq
where  $d$ is a constant. In flat coordinates $E$ is a linear vector 
field and the condition (\ref{qop}) becomes
$$\sum_{\alpha}\left(d_{\alpha}t^{\alpha}+r^{\alpha}\right)\f{\d F}{\d v^{\alpha}}=(3-d)F.$$
where $r^\alpha$ is a  constant, non--vanishing only if $d_{\alpha}=0$.

Any Frobenius manifold possesses a second flat metric defined, 
in flat coordinates for the first metric,
 by the formula
$$g^{\alpha\beta}=E^{\epsilon}c^{\alpha\beta}_{\epsilon}.$$

Using the Dubrovin-Novikov results, starting from the two flat metrics 
$\eta$ and $g$ one can define the following pair of Hamiltonian structures of 
hydrodynamic type:
\begin{eqnarray}
\label{P1} 
P_1^{\alpha\beta}&=&\eta^{\alpha\beta}\d_x\\
\label{P2}
P_2^{\alpha\beta}&=&g^{\alpha\beta}\d_x+\Gamma^{\alpha\beta}_{\gamma}u^{\gamma}_x=E^{\epsilon}c^{\alpha\beta}_{\epsilon}\d_x+\left(\f{d-1}{2}+d_{\beta}\right)c^{\alpha\beta}_{\gamma}.
\end{eqnarray}
It turns out \cite{du93} that $P_1$ and $P_2$ are compatible and 
therefore define a bi-Hamiltonian hierarchy of hydrodynamic type.
According to well-known results (see for instance \cite{cmp92}), the Hamiltonian densities of such a hierarchy
can be taken as the coefficients of the expansion at $\lambda=\infty$
\beq\label{casimirs}
c^{\alpha}(x,\lambda)= c^{\alpha}_{−1}(x)+\f{c^{\alpha}_0(x)}{\lambda}+
\f{c^{\alpha}_1(x)}{\lambda^2} + \dots  \qquad\lambda\to\infty,
\eeq
of the Casimirs of the pencil
$$P_2-\lambda P_1.$$
Since the Casimirs of a Hamiltonian structure of hydrodynamic type 
coincide with the flat coordinates of the corresponding metric,
 it follows that the Casimirs (\ref{casimirs}) are given by 
the flat coordinates of the pencil of metrics
\begin{equation}\label{fpom}
g-\lambda\eta,
\end{equation}
and thus satisfy the \emph{Gauss-Manin system}:
\beq\label{gms}
(\nabla^*-\lambda\nabla)d c^{\alpha}=0.
\eeq
Here $\nabla^*$ is the Levi-Civita connection for the metric $g$, and $\nabla$
 is the Levi-Civita connection for the metric $\eta$.

In this way, given a Frobenius manifold, it is possible to define a 
bi-Hamiltonian hierarchy  of hydrodynamic type. In flat coordinates for 
the metric $\eta$ the equations of such a hierarchy read 
\begin{equation}\label{biHh}
v^{\beta}_{t_{\alpha,k}}=
\eta^{\beta\gamma}\d_x\f{\delta H _{\alpha,k}}{\delta  v^{\gamma}},\qquad
\alpha,\beta=1,\dots,n,\quad k=-1,0,1,\dots
\end{equation}
where
$$H _{\alpha,k}=\int c_{\alpha,k}\,dx.$$

We now focus our attention on a special class of Frobenius manifolds. 

A Frobenius manifold $M$ is  called \emph{semisimple} \cite{du93} if at 
a generic point $v\in M$ the Frobenius algebra $T_v M$ is semisimple.
 The canonical coordinates
 $(u^1,\dots,u^n)$, whose existence is not an additional assumption but follows from the general  properties of these manifolds, can be obtained as solution of the equation
$${\rm det}(g-\lambda\eta)=0.$$

It turns out that, in canonical coordinates, the metrics $\eta$ and $g$ 
are diagonal and that the metric $\eta$ is of Egorov type. Moreover
such canonical coordinates are Riemann invariants of the hierarchy (\ref{biHh}).

Given a semisimple Frobenius manifold with $d<1$
it is possible to define a funcion $\lambda(p,u^1,\dots,u^n)$
called its \emph{superpotential}, having the following properties 
(for details see \cite{du93,du99,DZ}):

- it is defined as the inverse 
 of a special solution of the Gauss-Manin system (\ref{gms}) .

- its critical values  are the canonical coordinates.

- the covariant components of the metric $\eta$ in canonical 
coordinates can be written as
\begin{equation}\label{qesp}
\eta_{ij}=-\sum_{i=1}^n {\rm res}_{p=p_i}\frac{\d_i\lambda\d_j\lambda}{\lambda_p}\,dp=
-\frac{1}{2\pi i}\int_{\Gamma}\frac{\d_i\lambda\d_j\lambda}{\lambda_p}\,dp
\end{equation}
where $\Gamma$ are ``small'' closed contours around the critical points 
$p_1,\dots,p_n$ of $\lambda$.

Using these results it is easy to prove the following theorem
\begin{thm}
Let $M$ be a semisimple Frobenius manifold with $d<1$ and let 
$\lambda(p,u^1,\dots,u^n)$ be its superpotential, then the covariant 
and contravariant components of the metric $\eta$ in canonical 
coordinates admit the following quadratic expansions
\begin{eqnarray*}
&&\eta_{ij}=\f{1}{2\pi i}\int_C \d_i p(\lambda,u^1,\dots,u^n)\d_j p(\lambda,u^1,\dots,u^n)d\lambda\\
&&\eta^{ij}=\f{1}{2\pi i}\int_C W_i(\lambda,u^1,\dots,u^n)W_j(\lambda,u^1,\dots,u^n)d\lambda
\end{eqnarray*}
where $C$ are the images of the contour $\Gamma$ in the $\lambda$ plane and
 the functions $W^i(\lambda,u^1,\dots,u^n)$, defined by:
$$W^i(\lambda,u^1,\dots,u^n)=\f{\d_i p(\lambda,u^1,\dots,u^n)}{\eta_{ii}},$$
are generating functions of the symmetries of the semi-Hamiltonian hierarchy
 associated with $M$.
\end{thm}

\n
\emph{Proof.}  The first quadratic expansion can be obtained just by 
changing the variable $p\to\lambda$ in the integral (\ref{qesp}). 
Raising the indices we get the second quadratic expansion. 
In order to prove that the functions $W^i$ are generating functions of symmetries
it is sufficient to observe that the metric $\eta_{ii}$ is of Egorov 
type and that the inverse of the superpotential is a generating 
function of Hamiltonian densities. 

\begin{flushright}
$\Box$
\end{flushright}

\begin{rmk}
Starting from a Frobenius manifold one can define a hierarchy of 
integrable PDEs also in the following way. Let $\nabla$ be the Levi Civita connection associated with the metric $\eta$ and
 $(X_{(\alpha,0)},\alpha=1,\dots,n)$ be a basis of covariantly constant vector fields. One can define the \emph{primary flows} of the hierarchy as
$$u^i_{t_{(\alpha,0)}}=c^i_{jk}X^k_{(\alpha,0)}u^j_x,\,\,\,i=1,\dots,n.$$
and the ``higher flows''
\beq\label{ph}
u^i_{t_{(\alpha,n)}}=c^i_{jk}X^k_{(\alpha,n)}u^j_x,\,\,\,i=1,\dots,n,
\eeq
recursively, by means of the relations
$$\nabla_i X^k_{(\alpha,n)}=c^i_{kl}X^k_{(\alpha,n-1)}.$$
The hierarchy defined in this way is usually called the
\emph{principal hierachy}. 
It is equivalent to the hierachy defined above in 
terms of coefficients of the Casimirs of the pencil (\ref{fpom}) 
since the flows (\ref{ph}) are related to the flows 
(\ref{biHh}) just by triangular linear transformations 
(see \cite{du93,DZ} for details).
\end{rmk}
This shows that in the case of hierarchies of quasilinear 
 PDEs associated with a Frobenius manifold
the "factorization" (\ref{shtcW})
has a natural interpretation: the structure constants coincide 
with the structure constants defining the Frobenius structure, 
and the vector fields $X$ have a precise geometrical meaning.

\section{The classical shallow water equations}
Let us consider the classical shallow water system, given by
\begin{align}\label{sw}
h_t&=\left(hu\right)_x,\notag\\
&\\
u_t&=uu_x+h_x.\notag
\end{align}
A related problem was solved by Riemann, \cite{Rie}, 
using the hodograph transformation.

This system can be seen as the elementary $2-$component reduction 
of the Benney chain associated with the rational map 
 \cite{be73,za80}:
\beq\label{lambzakh}
\lm=p+\,\frac{h}{p-u},
\eeq
Moreover, (\ref{sw}) is an element of a bi-Hamiltonian hierarchy 
associated with a $2$ dimensional Frobenius manifold, with Frobenius potential
$$F(h,u)=\frac{1}{2}\,h u^2+ h \log{h},$$
in this setting, the function (\ref{lambzakh}) is the superpotential. 
Let us recall that under the change of coordinates
\begin{align*}
r^1=&\,u-2\,\sqrt{h},\\
r^2=&\,u+2\,\sqrt{h},
\end{align*}
the system takes the diagonal form
\begin{align*}
r^1_t=\frac{1}{4}\,\left(3\,r^1+\,r^2\right)r^1_x\\
r^2_t=\frac{1}{4}\,\left(r^1+3\,r^2\right)r^2_x.
\end{align*}
The general solution of the linear system (\ref{meq}), in this case, is
\begin{equation}\label{metzak}
g_{ii}=\varphi_i(r^i)\de_i A^0,\,\,\,i=1,2,
\end{equation}
where $\varphi_i(r^i)$ are arbitrary functions of a single variable and
$$A_0=\f{(r^1-r^2)^2}{16}.$$
We will show now that the quadratic expansion of the contravariant
components of the metrics (\ref{metzak}) can be reduced to a finite sum,
so that we can construct families of purely nonlocal Poisson brackets 
involving only a finite number of symmetries. We proceed as follows: 
first, we extend $\lm(p)$ from the upper half plane to a rational 
function on the whole Riemann sphere. Then, we note that although the 
extended $\lm$ is univalent, its inverse with respect to $p$ is not, 
and we have to consider two functions $p_+(\lm)$ and $p_-(\lm)$ each of 
them defined on one sheet of a double covering of the Riemann sphere, 
with branch points at $r^1$ and $r^2$. The two functions are easily 
found to be
\begin{align*}
p_+(\lm)=
\frac{1}{2}\,\lambda+\frac{1}{4}\,(r_1+\,r_2)+
\frac{1}{2}\,\sqrt {(r_2-\lambda)( r_1-\lambda) },\\
\\
p_-(\lm)=
\frac{1}{2}\,\lambda+\frac{1}{4}\,(r_1+\,r_2)-
\frac{1}{2}\,\sqrt {(r_2-\lambda)( r_1-\lambda) },
\end{align*}
their main difference being in the behaviour at infinity, for:
$$p_+(\lm)=\lm+O\left(\frac{1}{\lm}\right), \qquad \lm\rightarrow\infty_+,$$
while
$$p_-(\lm)=\f{u}{2}+O\left(\frac{1}{\lm}\right), \qquad \lm\rightarrow\infty_-.$$
Thus, we can construct two generating functions of the symmetries, the 
first is given by
$$w^i(\lm)=\frac{1}{p_+(\lm)-v^i},$$
whose expansion at infinity is
\beq\label{firstexp}
w^i(\lm)=\sum_{n=1}^\infty\frac{w_n^i}{\lm^n},
\eeq
and where the first few coefficients are given by
$$
\begin{tabular}{lll}
&$w_1^1=1\,$, &$w_1^2=1\,$,\\
\\
&$w_2^1=\frac{3}{4}\,r^1\!+\frac{1}{4}\,r^2$\,, &$w_2^2=\frac{1}{4}\,r^1\!+\frac{3}{4}\,r^2$,\\
\\
&$w_3^1=\frac{5}{8}\left(r^1\right)^2+\frac{1}{4}r^1r^2+\frac{1}{8}\left(r^2\right)^2$,
&$w_3^2=\frac{1}{8}\left(r^1\right)^2+\frac{1}{4}r^1r^2+\frac{5}{8}\left(r^2\right)^2$.\\
\end{tabular}
$$
For the second generating function, an easy calculation shows that the 
analogous generating function constructed from $p_-(\lm)$ is related 
with the first by
$$\frac{1}{p_-(\lm)-v^i}=w_0^i-\frac{1}{p_+(\lm)-v^i},\qquad i=1,2,$$
where
$$w_0^1=\!-\,\frac{4}{r^1-r^2}, \qquad w_0^2=\frac{4}{r^1-r^2},$$
is an extra symmetry, not appearing in the expansion (\ref{firstexp}).

Reducing the integral (\ref{qexpbr}) to the sum of residues around the 
two points  at infinity, $\infty_{+},\infty_{-}$, we obtain a finite 
quadratic expansion of the components of the metric tensor in terms of symmetries:
\begin{eqnarray*}
g^{ii}_{(k)}\,\delta^{ij}&=&\frac{(r^i)^k\delta_{ij}}{\de_i A^0}=
-\underset{\lm=\infty_+}{\rm Res}\frac{\lm^k\,d\lm}{(v^i-p_{+}(\lm))(v^j-p_{+}(\lm))}-
\underset{\lm=\infty_-}{\rm Res}\frac{\lm^k\,d\lm}{(v^i-p_{-}(\lm))(v^j-p_{-}(\lm))}\\
&=&\underset{\lm=\infty_{+}}{\rm Res}\frac{w_0^i\lm^k d\lm}{p_{+}(\lm)-v^j}
+\underset{\lm=\infty_{+}}{\rm Res}\frac{w_0^j\lm^k d\lm}{p_{+}(\lm)-v^i}-2\,
\underset{\lm=\infty_{+}}{\rm Res}\frac{\lm^k\,d\lm}{(v^i-p_{+}(\lm))(v^j-p_{+}(\lm))}
\\
&=&w_0^i\, w_{k+1}^j+w_{k+1}^i w_0^j-2\,\sum_{s=1}^k w^i_s\, w^j_{k-s+1}.
\end{eqnarray*}
Therefore, for $k=0,1,\dots ,$ the corresponding purely nonlocal 
Poisson operators have the form
$$P_{(k)}^{\,ij}=w_0^i\,\,r^i_x\left(\frac{d}{dx}\right)^{-1}\! \!w_{k+1}^j\,r^j_x+
w_{k+1}^i\,r^i_x\!\left(\frac{d}{dx}\right)^{-1}\!w_0^j\,r^j_x-2\,
\sum_{s=1}^k\left( w^i_s\, r^i_x\left(\frac{d}{dx}\right)^{-1}\!\!w^j_{k-s+1}\,r^j_x\right).$$

We consider now the flows generated by the simplest of these structures,
namely $P_{(0)}$. As Hamiltonian density we consider the generating 
function $p_+(\lm)$; the quantities we want to evaluate are thus the flows
$$\mu^i(\lm):=\sum_{j=1}^2P^{\,ij}_{(0)}\frac{\de p_+(\lm)}{\de \lm^j}.$$
Explicitly, and assuming vanishing  boundary conditions 
$$\lim_{|\,x|\to\infty}r^i(x,t)=0,$$
these are found to be
\begin{align*}
\mu^1(\lm)&=-4\,\,\frac{p_+(\lm)-\lm}{r^1-r^2}-2\,\ln{\left(\sqrt{\lm-r_1}+\sqrt{\lm-r_2}\right)}+\ln{4\lm},\\
\mu^2(\lm)&=4\,\,\frac{p_+(\lm)-\lm}{r^1-r^2}-2\,\ln{\left(\sqrt{\lm-r_1}+\sqrt{\lm-r_2}\right)}+\ln{4\lm}.
\end{align*}
By comparing the coefficients of the asymptotic expansions at infinity 
of $\mu^i(\lm)$ and $p_+(\lm)$ we obtain, for instance, that the 
characteristic velocities $w_2$  of the systems are generated by the 
Hamiltonian density $H^0$, while the symmetry $w_3$ is obtained from 
$H^1$. For the shallow water hierarchy, there exist another Poisson 
structure which sends the Hamiltonian density $H^0$ to the system with 
characteristic velocities $w_2$. This is the third {\em local} Poisson 
structure of the system, generated by the flat metric 
$g_{ii}=\frac{2}{\left(\lm^i\right)^2}\de_i A^0$. 
We denote this structure as $P_{loc}$, and we call $z^i(\lm)$ 
the corresponding flows, so that
$$z^i(\lm)=\sum_{j=1}^2P_{loc}^{\,ij}\frac{\de p_+(\lm)}{\de \lm^j}.$$
With little difficulty, it can be shown that the two Hamiltonian 
hierarchies $z^i(\lm)$ and $\mu^i(\lm)$ are related by
$$z^i(\lm)=\frac{\lm^2}{2}\frac{d^{\,2}\mu^i(\lm)}{d\,\lm^2},$$
so that the coefficients of the expansion at infinity are related by 
$$z^i_k=\frac{k(k+1)}{2}\,\mu^i_k.$$
Moreover, the original generating function of the symmetries $w^i(\lm)$ 
can written in terms of $\mu^i(\lm)$ as
$$w^i(\lm)=\frac{1}{\lm}-\frac{d\,\mu^i(\lm)}{d\,\lm},$$
so that the coefficients satisfy the relation
$$w^i_{k+1}=-(k+1)\,\mu^i_{k}.$$

\section*{Acknowledgments}
We would like to thank the ESF grant MISGAM 1414, for its support of Paolo Lorenzoni's visit to Imperial.
We are also grateful to the European Commission�s FP6 programme for support of this work through the ENIGMA network,
and particularly for their support of Andrea Raimondo, who also received an EPSRC DTA.

\end{document}